\global\def\draftcontrol{0}

   \def\versionno{ n=2 critical }

\catcode`\@=11

\expandafter\ifx\csname draftcontrol\endcsname\relax\global\def\draftcontrol{0}
\fi

{\count255=\time\divide\count255 by 60
\xdef\hourmin{\number\count255}
\multiply\count255 by-60\advance\count255 by\time
\xdef\hourmin{\hourmin:\ifnum\count255<10 0\fi\the\count255}}
\def\draftdate{\number\month/\number\day/\number\year\ \ \ \hourmin }

\newcommand\makepapertitle{\par
  \begingroup
    \renewcommand\thefootnote{\@fnsymbol\c@footnote}%
    \def\@makefnmark{\rlap{\@textsuperscript{\normalfont\@thefnmark}}}%
    \long\def\@makefntext##1{\parindent 1em\noindent
            \hb@xt@1.8em{%
                \hss\@textsuperscript{\normalfont\@thefnmark}}##1}%
     \newpage
     \global\@topnum\z@   
     \@makepapertitle
     \thispagestyle{empty}\@thanks
  \endgroup
  \setcounter{footnote}{0}%
  \global\let\thanks\relax
  \global\let\makepapertitle\relax
  \global\let\@makepapertitle\relax
  \global\let\@thanks\@empty
  \global\let\@author\@empty
  \global\let\@date\@empty
  \global\let\@title\@empty
  \global\let\title\relax
  \global\let\author\relax
  \global\let\date\relax
  \global\let\and\relax
  \def\version{\let\version\@version\@gobble}
}
\def\@makepapertitle{%
  \newpage
   \ifnum\draftcontrol=1 {}
   \version\versionno
   \vskip 3em%
   \else
   \hfill\hbox to 3cm {\parbox{4cm}{\@pubnum}\hss}%
   \vskip 3em%
   \fi
   \begin{center}%
   \let \footnote \thanks
     {\LARGE {\@title}}%
     \vskip 1.5em%
     {\normalsize
       \lineskip .5em%
       \begin{tabular}[t]{c}%
         \@author
       \end{tabular}\par}%
     \vskip 1.5em%
     {\@bstract}%
     \end{center}%
     \vskip 1.5em
     \@date%
   \par
}

\gdef\@pubnum{}
\def\pubnum#1{%
  \gdef\@pubnum{#1}}

\gdef\@bstract{}
\def\Abstract#1{%
  \gdef\@bstract{%
   \parbox{\textwidth-0pc}{%
   \centerline{\bf Abstract}\penalty1000%
\kern.2cm%
\noindent
\renewcommand\baselinestretch{1.0}%
{#1}}}
}

\def\ps@paper{\let\@mkboth\@gobbletwo%
     \ifnum\draftcontrol=1
    \def\@oddfoot{\hbox to \textwidth{\tiny \versionno \hfil\tiny\draftdate}%
    \hskip -\textwidth \hbox to \textwidth{\hfil\rm\thepage\hfil}}%
     \else\def\@oddfoot{\hbox to \textwidth{\hfil\rm\thepage\hfil}}
     \fi
     \let\@evenfoot\@oddfoot
}

\def\body{\clearpage
          \pagestyle{paper}
    }

\def\@version#1{\ifnum\draftcontrol=1
\typeout{}\typeout{#1}\typeout{}
\vskip3mm\centerline{\hbox{\fbox{\normalsize{\tt DRAFT -- #1 -- }
                   {\draftdate}}}}\vskip3mm
\fi}
\let\version\@version
\long\def\eqlabel#1{\ifnum\draftcontrol=1
                    \tag@false  
                    \tag*{(\theequation) \hbox to -0.2cm{\hspace{0cm}\small{#1}\hss}}
                    \refstepcounter{equation}
                    \edef\@currentlabel{\theequation}
                    \ltx@label{#1}          
                    \else
                    \label{#1}
                    \fi
                    }
\let\st@bibitem\@bibitem
\let\st@lbibitem\@lbibitem
\ifnum\draftcontrol=1
  \def\@bibitem#1{%
    \st@bibitem{#1}\a@@label{#1}\ignorespaces}
  \def\@lbibitem[#1]#2{%
    \st@lbibitem[#1]{#2}\a@@label{#2}\ignorespaces}
  \def\a@@label#1{%
    \gdef\a@lab{\smash{\normalfont\small#1}}
    \ifvmode
      \if@inlabel
        \global\setbox\@labels\hbox{%
          \llap{\a@lab\let\a@lab\relax
                \kern\@totalleftmargin\kern\marginparsep}%
          \box\@labels}%
      \fi
    \fi}
\fi

\documentclass[12pt,letterpaper]{article}

\usepackage{amsmath,amssymb,array,calc,epsfig,rotating,bm}
\usepackage[sort]{cite}
\usepackage{graphicx}
\usepackage{psfrag,verbatim}

\ifnum\draftcontrol=1
\fi

\renewcommand\baselinestretch{1.25}
\renewcommand\section{\@startsection {section}{1}{\z@}%
                                   {-3.5ex \@plus -1ex \@minus -.2ex}%
                                   {2.3ex \@plus.2ex}%
                                   {\normalfont\large\bfseries}}
\renewcommand\subsection{\@startsection{subsection}{2}{\z@}%
                                   {-3.25ex\@plus -1ex \@minus -.2ex}%
                                   {1.5ex \@plus .2ex}%
                                   {\normalfont\normalsize\bfseries}}
\renewcommand\subsubsection{\@startsection{subsubsection}{3}{\z@}%
                                   {-3.25ex\@plus -1ex \@minus -.2ex}%
                                   {1.5ex \@plus .2ex}%
                                   {\normalfont\normalsize\it}}
\renewcommand\paragraph{\@startsection{paragraph}{4}{\z@}%
                                   {-3.25ex\@plus -1ex \@minus -.2ex}%
                                   {1.5ex \@plus .2ex}%
                                   {\normalfont\normalsize\bf}}

\numberwithin{equation}{section}

\def\revise#1       {\raisebox{-0em}{\rule{3pt}{1em}}%
                     \marginpar{\raisebox{.5em}{\vrule width3pt\
                     \vrule width0pt height 0pt depth0.5em
                     \hbox to 0cm{\hspace{0cm}{%
                     \parbox[t]{4em}{\raggedright\footnotesize{#1}}}\hss}}}}

\newcommand\nxt[1]  {\\\fnxt#1}

\def\calc         {{\cal C}}

\def\cale         {{\cal E}}
\def\calf         {{\cal F}}

\def\calh         {{\cal H}}

\def\call         {{\cal L}}
\def\calm         {{\cal M}}
\def\caln         {{\cal N}}
\def\calo         {{\cal O}}
\def\calp         {{\cal P}}

\def\calw         {{\cal W}}
\def\calz         {{\cal Z}}

\def\del          {\partial}

\def\tr           {\mathop{\rm Tr}}

\def\sqr#1#2{{\vcenter{\vbox{\hrule height.#2pt
 \hbox{\vrule width.#2pt height#1pt \kern#1pt
 \vrule width.#2pt}\hrule height.#2pt}}}}

\newcommand{\ft}[2]{{\textstyle{\frac{#1}{#2}}}}

\def\a{\alpha}
\def\b{\beta}
\newcommand{\qq}{\mathfrak{q}}
\newcommand{\ww}{\mathfrak{w}}

\newcommand{\beq}{\begin{equation}}
\newcommand{\eeq}{\end{equation}}
\newcommand{\beqa}{\begin{eqnarray}}
\newcommand{\eeqa}{\end{eqnarray}}
\newcommand{\beqar}{\begin{eqnarray*}}
\newcommand{\eeqar}{\end{eqnarray*}}

\renewcommand{\eqref}[1]{(\ref{#1})}

\newcommand{\ie}{{\it i.e.,}\ }

\def\a{\alpha}

\def\t{\tau}

\def\r{\rho}

\def\dd{{\delta}}
\def\l{\lambda}

\def\ra{\Rightarrow}
\def\c{\chi}

\catcode`\@=12

\begin{document}


\title{\bf Critical phenomena in $\caln=2^*$ plasma}
\pubnum
{UWO-TH-10/7
}

\date{October 2010}

\author{
Alex Buchel$ ^{1,2}$ and Chris Pagnutti$ ^{1}$\\[0.4cm]
\it $ ^1$Department of Applied Mathematics\\
\it University of Western Ontario\\
\it London, Ontario N6A 5B7, Canada\\
\it $ ^2$Perimeter Institute for Theoretical Physics\\
\it Waterloo, Ontario N2J 2W9, Canada\\
}

\Abstract{
We use gauge theory/string theory correspondence to study finite
temperature critical behavior of mass deformed $\caln=4$ $SU(N)$
supersymmetric Yang-Mills theory at strong coupling, also known as
$\caln=2^*$ gauge theory.  For certain range of the mass parameters,
$\caln=2^*$ plasma undergoes a second-order phase transition. We
compute all the static critical exponents of the model and demonstrate
that the transition is of the mean-field theory type. We show that the
dynamical critical exponent of the model is $z=0$, with multiple
hydrodynamic relaxation rates at criticality. We point out that the
dynamical critical phenomena in $\caln=2^*$ plasma is outside the
dynamical universality classes established by Hohenberg and Halperin.
}

\makepapertitle

\body

\version\versionno
\tableofcontents
\section{Introduction}
Gauge theory/string theory correspondence \cite{m9711} presents a solvable framework 
to study a large class of strongly interacting four-dimensional gauge theory plasmas.
In a nutshell, the  solvability of these models comes from ability to approximate 
a dual  string theory with a corresponding classical supergravity.   
Unfortunately, real QCD is not any one of the models studied. It is possible to 
reach QCD as a particular limit in some of these models, but in doing so the truncation 
of the string theory to a supergravity sector becomes inconsistent. Instead 
one attempts to discover common/universal features of strongly coupled gauge theory 
plasmas, and hopes that real QCD is in the universality class of the models studied. 
Typical examples of such universal properties are the 
strongly coupled plasma shear and bulk viscosities:
\nxt the shear viscosity $\eta$ to entropy density $s$ ratio \cite{u1,u2,u3,u4}
\begin{equation}
\frac \eta s =\frac{1}{4\pi}\,,
\eqlabel{etas}
\end{equation}
\nxt the bulk viscosity bound \cite{bulk1}
\begin{equation}
\frac \zeta \eta\ge 2\left(\frac 13-c_s^2\right)\,,\qquad c_s^2=\frac{\del \calp}{\del\cale}\,.
\eqlabel{zetaeta}
\end{equation}
Although above properties of strongly coupled plasmas have been observed 
(or in case of the shear viscosity derived) in holographic setting, it is  
not clear why and how these universalities arise, or how to properly 
{\it define} the corresponding universality class: while the
shear viscosity ratio in universal in 2-derivative 
supergravity\footnote{This translates into an infinite t' Hooft coupling limit on 
the gauge theory side.}  (or a phenomenological model of thereof), it can 
be violated in full string theory \cite{v1,v2,v3}; while the bulk viscosity bound is 
satisfied in all models of supergravity derived from string theory, it can 
be violated in some phenomenological models of gauge/gravity correspondence 
\cite{gpr}. 

A more common notion of the 'universality' arises in the theory of continuous critical 
phenomena. In this paper we follow up the work of \cite{cc1,cc2,cc3} and focus on static 
and dynamic properties of strongly coupled non-conformal gauge theory plasma in the vicinity 
of the second-order phase transition. In \cite{cc3} it was noticed that there was a 
tension between the hyperscaling relation among the static critical exponents 
at the second-order phase transition, and the expectation that in the planar limit 
the transition should be of the mean-field type, \ie with vanishing anomalous 
critical exponent\footnote{Contrary to some statements in 
recent literature (as in \cite{franco} for instance), we take a perspective 
here that for a second-order phase transition to be of a 
mean-field type the anomalous critical exponent must vanish --- whether or not 
the other critical exponents are integers or not is irrelevant. }. 
Direct computation of  critical exponents for the second-order 
phase transition in $\caln=4$ SYM plasma at finite temperature and the chemical 
potential for a global $U(1)_R$ charge confirmed the vanishing of the anomalous
critical exponent. Further, the dynamical critical exponent of this 
transition was shown to be $z=4$, even though the background geometry 
at criticality did not exhibit a $z=4$ Lifshitz-like scaling. In other words,
the transition detailed in \cite{cc3} explicitly showed that the dynamical 
scaling properties can be ``emergent'' and should not be necessarily ``enforced''
on the background geometry of the holographic dual.  

Although we restrict our attention here to a second-order phase transition in 
mass-deformed $\caln=4$ SYM 
(also known as $\caln=2^*$ gauge theory \cite{pw,bpp,ejp}), we emphasize that 
the holographic (static) universality class of this transition includes also 
a cascading gauge theory \cite{cca}. In section 2 we review the holographic 
duality for $\caln=2^*$ gauge theory plasma. Critical phenomena in $\caln=2^*$
plasma from both the gauge theory and the dual gravitational perspective is discussed in 
section 3.  Some of the static critical exponents
  of the second-order phase transition in this plasma,
namely $\{\a,\b,\gamma,\delta\}$, were computed 
in \cite{cc2}. We directly compute the remaining static critical exponents $\{\nu,\eta\}$ 
and the dynamical critical exponent $z$ of the theory in section 4. 
We collect all the results in section 5.

\section{$\caln=2^*$/PW holographic duality}

In this section we briefly review the main features of the holographic duality 
between $\caln=2^*$ $SU(N)$ gauge theory and the Pilch-Warner (PW) 
geometry of type IIB 
supergravity. We refer the reader to the original work for further details 
\cite{pw,bpp,ejp,bl,n2hydro,bdkl,bp}.

Consider maximally supersymmetric $\caln=4$ $SU(N)$ Yang-Mills theory 
in the planar limit ($g_{YM}^2\to 0$, $N\to \infty$ with $\l\equiv g_{YM}^2 N$ kept fixed)
and for large 't Hooft coupling $\l\gg 1$. According to Maldacena correspondence 
\cite{m9711} this superconformal theory is equivalent to a classical type IIB supergravity 
on $AdS_5\times S^5$. A duality between a SYM  and a supergravity 
can be extended (on both sides) away from the conformal point \cite{pw,bpp,ejp}. 
On the gauge theory side, a massive deformation of $\caln=4$ superpotential 
\begin{equation}
W_{\caln=4}=\frac{2\sqrt{2}}{g_{YM}^2}\ \tr\left(\left[Q,\tilde{Q}\right]\Phi\right)\,,
\eqlabel{w4}
\end{equation}  
where $\{Q,\tilde{Q}, \Phi\}$ are $\caln=1$ adjoint chiral superfields,   
to 
\begin{equation}
W_{\caln=4}\to 
W_{\caln=2^*}=W_{\caln=4}
+\frac{m}{g_{YM}^2}\left(\tr Q^2+\tr \tilde{Q}^2\right)\,,
\eqlabel{w2}
\end{equation}
breaks half of the supersymmetries. This mass-deformed theory is known as 
$\caln=2^*$ gauge theory. When $m\ne 0$, the mass deformation lifts the 
$\{Q,\tilde{Q}\}$ $\caln=2$ hypermultiplet moduli directions, resulting in 
$(N-1)$ complex dimensional Coulomb branch parametrized by 
\begin{equation}
\Phi={\rm diag}\left(a_1,a_2,\cdots,a_N\right)\,,\qquad \sum_{i=1}^N\ a_i=0\,.
\eqlabel{cbranch}
\end{equation}
We study $\caln=2^*$ gauge theory at a particular point on the Coulomb branch moduli 
space \cite{bpp}:
\begin{equation}
a_i\in [-a_0,a_0]\,,\qquad a_0^2=\frac{m^2 g_{YM}^2 N}{\pi}\,,
\eqlabel{pointc}
\end{equation}
with the (continuous in the large $N$-limit) linear number density 
\begin{equation}
\r(a)=\frac{2}{m^2 g_{YM}^2}\sqrt{a_0^2-a^2}\,,\qquad \int_{-a_0}^{a_0}\ da\ \r(a)=N\,.
\eqlabel{distr}
\end{equation}
The reason for such an esoteric choice for a vacuum of the theory is simply because we know 
a dual holographic description of the theory (as a Pilch-Warner geometry \cite{pw}) only at this 
point \cite{bpp}. Extending the correspondence to the rest of the moduli space is an important 
unsolved problem.   

Notice that the deformation  \eqref{w2} is actually a deformation of a CFT by two different operators: 
a dimension-2 operator (a mass term for the bosonic components of the $\{Q,\tilde{Q}\}$
hypermultiplet) and a dimension-3 operator (a mass term for the fermionic components of the 
$\{Q,\tilde{Q}\}$ hypermultiplet). According to AdS/CFT dictionary \cite{mr}, a scalar gauge-invariant operator 
of dimension $\Delta$ is dual to a scalar field of mass
$m_5^2 L^2=\Delta(\Delta-4)$ of the five-dimensional dual gravitational description.  These two mass-deformation  
operators are the $\a$ and $\chi$ scalars of the Pilch-Warner effective action \cite{pw}:
\begin{equation}
\begin{split}
S=
\int_{\calm_5} d\xi^5 \sqrt{-g}\ \call_5
=\frac{1}{4\pi G_5}\,
\int_{\calm_5} d\xi^5 \sqrt{-g}\left[\ft14 R-3 (\del\a)^2-(\del\chi)^2-
\calp\right]\,,
\end{split}
\eqlabel{action5}
\end{equation}
where the potential%
\footnote{We set the five-dimensional gauged
supergravity coupling to one. This corresponds to setting the
radius $L$ of the five-dimensional sphere in the undeformed metric
to $2$.}
\begin{equation}
\calp=\frac{1}{16}\left[\frac 13 \left(\frac{\del W}{\del
\a}\right)^2+ \left(\frac{\del W}{\del \chi}\right)^2\right]-\frac
13 W^2\,,
 \eqlabel{pp}
\end{equation}
is a function of $\alpha$ and $\chi$, and is determined by the
superpotential
\begin{equation}
W=- e^{-2\alpha} - \frac{1}{2} e^{4\alpha} \cosh(2\chi)\,.
\eqlabel{supp}
\end{equation}
In our conventions, the five-dimensional Newton's constant is
\begin{equation}
G_5\equiv \frac{G_{10}}{2^5\ {\rm vol}_{S^5}}=\frac{4\pi}{N^2}\,.
\eqlabel{g5}
\end{equation}

In what follows we focus on equilibrium thermal states of $\caln=2^*$ plasma. 
Their  holographic dual is represented by a regular black brane solution in the 
effective action \eqref{action5} \cite{bl,bdkl}
\begin{equation}
ds_5^2=e^{2A}\left(-(1-x)^2 dt^2+dx_1^2+dx_2^2+dx_3^2\right)+g_{xx}dx^2\,,
\eqlabel{background}
\end{equation} 
with $g_{xx}=g_{xx}(x)$, $A=A(x)$, $\a=\a(x)$ and $\chi=\chi(x)$ being  functions of the 
radial coordinate $x\in [0,1]$ only. Note that $x\to 0_+$ corresponds to the asymptotic $AdS_5$
boundary, while $x\to 1_-$ to a regular Schwarzschild horizon. The temperature and the 
mass parameters of the plasma are encoded in the asymptotic behavior of the 
supergravity fields $\{A,\a,\chi\}$. Specifically, near the $AdS_5$ boundary we have 
\begin{equation}
e^\a\equiv \r=1+x^{1/2}\left(\r_{10}+\r_{11}\ \ln x \right)+\cdots+x^{k/2}\left(
\sum_{i=1}^k\ \r_{ki}\ \ln^i x
\right)+\cdots\,,
\eqlabel{rb}
\end{equation} 
\begin{equation}
\chi=\chi_0 x^{1/4}\left[1+x^{1/2}\left(\chi_{10}+\chi_{11}\ \ln x\right)
+\cdots+x^{k/2}\left(
\sum_{i=1}^k\ \chi_{ki}\ \ln^i x
\right)+\cdots
\right]\,,
\eqlabel{cb}
\end{equation} 
\begin{equation}
a=x^{1/2}\left(a_{10}+a_{11}\ \ln x\right)
+\cdots+x^{k/2}\left(
\sum_{i=1}^k\ a_{ki}\ \ln^i x
\right)+\cdots\,,
\eqlabel{ab}
\end{equation} 
and 
\begin{equation}
\r=\r_h+\r_1 (1-x)^2+\cdots+\r_k (1-x)^{2k}+\cdots\,,
\eqlabel{rh}
\end{equation}
\begin{equation}
\chi=\chi_h+\chi_1 (1-x)^2+\cdots+\chi_k (1-x)^{2k}+\cdots\,,
\eqlabel{ch}
\end{equation}
\begin{equation}
a=a_h+a_1 (1-x)^2+\cdots+a_k (1-x)^{2k}+\cdots\,,
\eqlabel{ah}
\end{equation}
near the regular Schwarzschild horizon. In \eqref{ab}, \eqref{ah} $a(x)$  is defined as 
\begin{equation}
A(x)\equiv \ln\hat\delta_3-\frac 14\ln(2x-x^2)+a(x)\,.
\eqlabel{defa}
\end{equation}
In was shown in \cite{bdkl} that given $\{\hat{\delta}_3,\r_{11},\chi_0\}$, there is a 
unique singularity-free solution of \eqref{action5} representing the equilibrium 
state of $\caln=2^*$ plasma. On the gravity side, the coefficients of the leading 
asymptotics, namely $\{\hat{\delta}_3,\r_{11},\chi_0\}$, determine 
the remaining 6 parameters (2 in the UV and 4 in the IR) of the solution: 
\begin{equation}
\begin{split}
&{\rm\ UV}:\ \{\r_{10},\,\chi_{10}\}\,,\\
&{\rm\ IR}:\ \{\r_h,\,\chi_{h},\, a_h,\, a_1\}\,.
\end{split}
\eqlabel{uvirp}
\end{equation}   
It is possible to unambiguously relate the gravitational and the gauge theory 
data \cite{bdkl}:
\nxt the plasma temperature $T$, and the masses $\{m_b,\, m_f\}$ of the bosonic 
and the fermionic components of the $\caln=2$ hypermultiplet are given 
by
\begin{equation}
T=\frac{\hat{\delta_3}}{2\pi }e^{-3 a_h}\,,\qquad \frac{m_b^2}{T^2}=12\sqrt{2}\pi^2 \r_{11} 
e^{6 a_h}\,,\qquad \frac{m_f}{T}=2^{3/4}\pi \chi_0 e^{3a_h}\,, 
\eqlabel{micro}
\end{equation}
\nxt and the plasma free energy density $\calf$,
the energy density $\cale$, and the entropy density $s$ are given by 
\begin{equation}
\begin{split}
&\calf=-\frac{\hat{\delta}_3^4}{32\pi G_5}\biggl(
1+\r_{11}^2\left(24-96\ln\hat{\delta}_3+24\ln 2\right)
-24\r_{10}\r_{11}+2\chi_0^2\chi_{10}\\
&+\chi_0^4\left(\frac 49-\frac 23\ln 2+\frac 83\ln\hat{\delta}_3\right)
\biggr)\,,\\
&\cale=\calf-\frac{1}{8\pi G_5}\hat{\delta}_3^4\,,\qquad 
s=\frac{\hat{\delta}_3^3e^{3a_h}}{4 G_5}\,.
\end{split}
\eqlabel{thermo}
\end{equation}
To recover the $\caln=2$ supersymmetric PW vacuum \eqref{pointc}, \eqref{distr}
we need to set $T=0$ and fine-tune the masses $m_b=m_f=m$.

\section{Critical phenomena in $\caln=2^*$ plasma}

\begin{figure}[t]
\begin{center}
\psfrag{cs2}{{$c_s^2$}}
\psfrag{r11}{{$\r_{11}$}}
\psfrag{mbT}{{$\frac{m_b}{T}$}}
\includegraphics[width=3in]{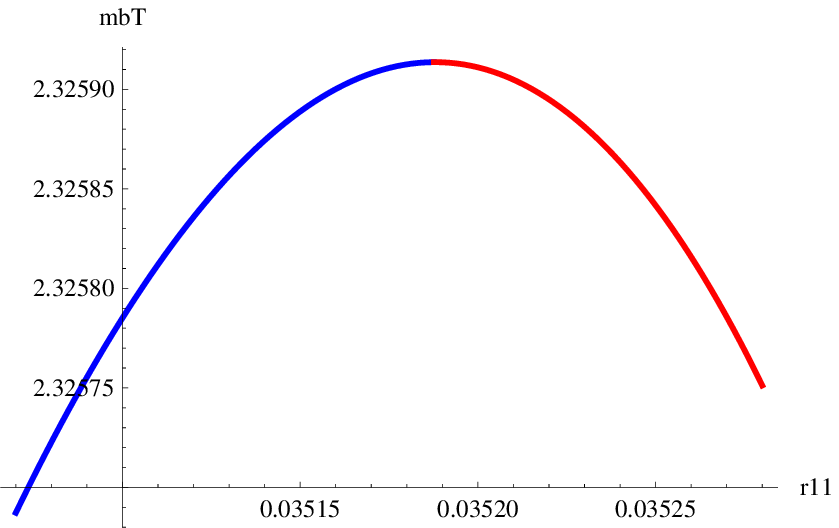}
  \includegraphics[width=3in]{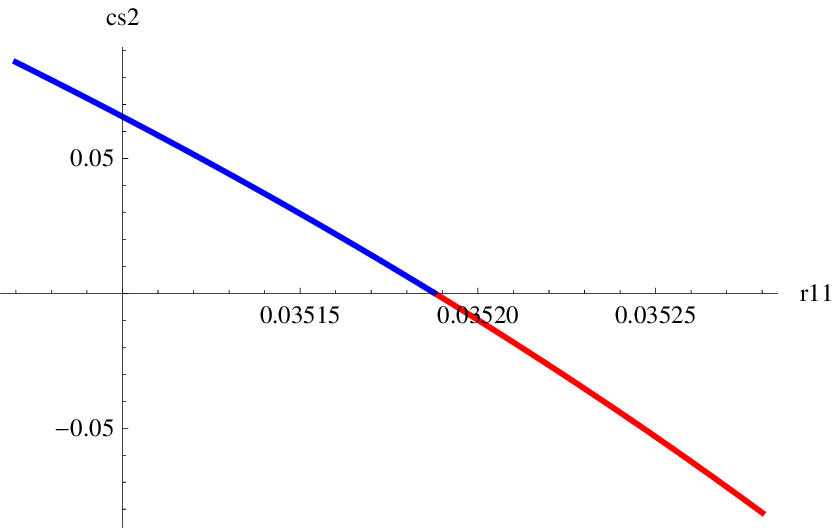}
\end{center}
  \caption{(Colour online)
The  dimensionless temperature $\frac{m_b}{T}$ (left plot) and the speed of sound $c_s^2$ (right plot)
of the strongly coupled $\caln=2^*$ plasma with $m_f=0$ and $m_b\ne 0$
as a function of the dual gravitation parameter $\r_{11}$.} \label{figure1}
\end{figure}

\begin{figure}[t]
\begin{center}
\psfrag{fred}{{$\Omega/\left(\frac 18 \pi^2 N^2 T^4\right)$}}
\psfrag{r11}{{$\r_{11}$}}
\psfrag{mbT}{{$\frac{m_b}{T}$}}
  \includegraphics[width=3in]{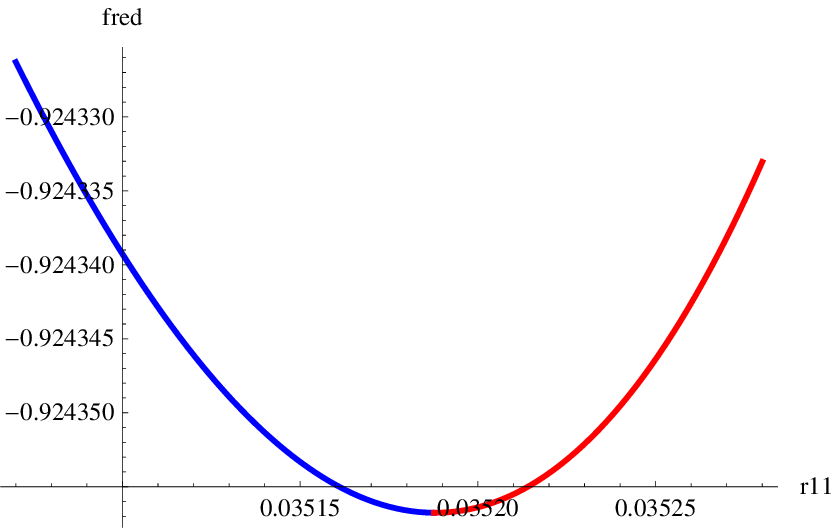}
\includegraphics[width=3in]{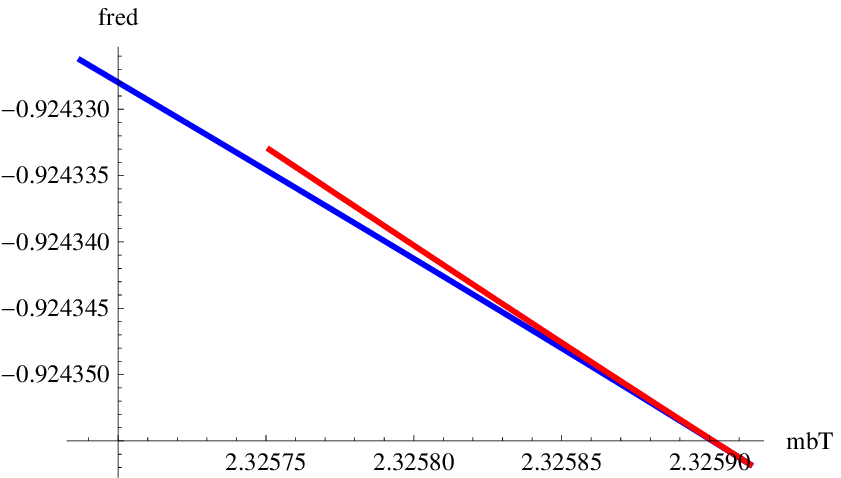}
\end{center}
  \caption{(Colour online)
Free energy densities $\Omega_o$ of the ``ordered'' phase (blue curves) 
and $\Omega_d$ of the ``disordered'' phase (red curves) as a function of 
$\r_{11}$ (left plot) and $\frac{m_b}{T}$ (right plot) of  $\caln=2^*$
plasma with $m_f=0$.} \label{figure2}
\end{figure}

The phase diagram of $\caln=2^*$ plasma was studied in details in \cite{bdkl,bp}. 
It was found there that whenever $m_f^2<m_b^2$, the theory undergoes a second-order phase transition. 
with the  critical temperature $T_c=T_c\left(\frac{m_f^2}{m_b^2}\right)$. All these transitions 
are in the same universality class, and thus we can restrict our attention to $m_f=0$, $m_b\ne 0$ case:
\begin{equation}
m_f=0:\qquad \frac{m_b}{T_c}\approx 2.32591\,.
\eqlabel{mftc}
\end{equation} 
We now recall the main characteristics of this transition \cite{cc2}:
\nxt
The left plot on Figure \ref{figure1} represents the dependence of the dimensionless temperature $\frac{m_b}{T}$
on the gravitational parameter $\r_{11}$, see \eqref{micro}. The transition is associated with the 
minimal accessible temperature, to be identified with $T_c$, in the plasma for isotropic and homogeneous 
equilibrium state\footnote{This feature of the transition is also observed for the phase transition in  
$\caln=4$ SYM plasma with a single $U(1)\subset SU(4)_R$ R-symmetry chemical potential \cite{cc3}.} .
For each temperature $T>T_c$ there are two phases --- the "ordered'' phase (blue curves), and the "disordered'' phase 
(red curves). The right plot on Figure \ref{figure1} represents the square of the speed of sound $c_s^2$ as a function 
of $\r_{11}$.  Notice that the hydrodynamic modes in the  "disordered phase'' are unstable (as $c_s^2<0$), and thus must condense.  
It is tempting to conjecture that the equilibrium state in the plasma at $T< T_c$ breaks translational
invariance, and represents the end point of condensation of hydrodynamic modes \cite{abh}. 
\nxt The free energy densities of the stable ``ordered'' phase (blue curves) $\Omega_o$ and the unstable "disordered'' 
phase (red curves) $\Omega_d$ as a function of the gravitational parameter $\r_{11}$ (left plot) and the dimensionless 
temperature $\frac{m_b}{T}$ (right plot) are represented in Figure \ref{figure2}. 
\nxt It is convenient to recast the critical behavior in $\caln=2^*$ plasma in that of a 3-dimensional ferromagnet. 
The thermodynamics of the latter is described by  the Gibbs free energy $\calw=\calw(t,\calh)$  which depends on the 
reduced temperature $t=(T-T_c)/T_c$ and  the external magnetic field  $\calh$. Once we identify
\begin{equation}
\calw\equiv \Omega_o-\Omega_d\,,\qquad \calh\equiv m_b\,,
\eqlabel{map}
\end{equation} 
and introduce 
\begin{equation}
\Delta\r_{11}\equiv \r_{11}-\r_{11}^c\,,\qquad |\Delta\r_{11}|\propto t^{1/2}\,,\qquad \r_{11}^{c}=0.035187(6)\,,
\eqlabel{derdr}
\end{equation}
we can compute the standard static critical exponents $\{\a,\b,\gamma,\delta\}$:
\begin{equation}
c_\calh=-T\left(\frac{\del^2\calw}{\del T^2}\right)_\calh\propto |t|^{-\a}=\frac{s}{c_s^2}\bigg|^{blue}_{red}\propto c_s^{-2}\bigg|^{blue}_{red}
\propto (\Delta\r_{11})^{-1}\propto t^{-1/2}\ \ \ra\ \ \a=\frac 12\,,
\eqlabel{alpha}
\end{equation}
\begin{equation}
\begin{split}
\calm=&-\left(\frac{\del\calw}{\del\calh}\right)_T\propto |t|^\beta\propto 
-\frac{1}{\Delta\r_{11}}{\del_{\Delta\r_{11}}\calw}\propto -\frac{1}{\Delta\r_{11}}\del_{\Delta\r_{11}}\ \left(-|\Delta\r_{11}|^3\right)
\\
&\propto -|\Delta\r_{11}|\propto -t^{1/2}\qquad \ra\qquad \b=\frac 12\,,
\end{split}
\eqlabel{beta}
\end{equation}
\begin{equation}
\chi_T=\left(\frac{\del\calm}{\del\calh}\right)_T\propto |t|^{-\gamma}\propto -\del_t \calm\propto \del_t t^{1/2}\propto t^{-1/2}
\qquad \ra\qquad \gamma=\frac 12\,,
\eqlabel{gamma}
\end{equation}
\begin{equation}
\calm(t=0)\propto |\calh-\calh_c|^{1/\delta}\propto t^{1/\delta}\propto t^{1/2}\qquad \ra\qquad \delta=2\,.
\eqlabel{delta}
\end{equation}
Thus,
\begin{equation}
\{\a,\b,\gamma,\delta\}=\left\{\frac 12,\frac 12,\frac 12, 2\right\}\,.
\eqlabel{sumcrit}
\end{equation}

The remaining two critical exponents $\{\nu,\eta\}$ are more difficult to extract as they are related to the 
scaling properties of the magnetization two-point correlation function at criticality:
\begin{equation}
G(\vec{r})=\langle\calm(\vec{r})\calm(\vec{0})\rangle\propto \frac{\del^2\calw}{\del\calh(\vec{r})\del\calh(\vec{0})}\,,
\eqlabel{cor1}
\end{equation}
\begin{equation}
G(\vec{r})\propto
\begin{cases}
&e^{-|\vec{r}|/\xi}\,,\qquad  t\ne 0\cr
&|\vec{r}|^{-3+2-\eta}\,,\qquad  t=0
\end{cases}\,,\qquad {\rm with}\qquad \xi\propto |t|^{-\nu} \,,
\eqlabel{defs01}
\end{equation}
where $\xi$ is the correlation length. Under the static scaling hypothesis, 
\begin{equation}
\calw(t,\calh)=\l^{-3}\ \calw(\l^{y_T}t,\l^{y_\calh}\calh)\,,\qquad \tilde{G}(\vec{q},t,\calh)=\l^{2 y_{\calh}-3}
\tilde{G}(\l\vec{q},\l^{y_T}t,\l^{y_\calh}\calh)\,,
\eqlabel{stsc1}
\end{equation}
where $y_{T}$ and $y_{\calh}$ are the two independent critical exponents; 
$\tilde G$ is a spatial Fourier transform of \eqref{cor1}. The static scaling hypothesis implies 
4 scaling relations between $\{\a,\b,\gamma,\delta,\nu,\eta\}$. In particular, using two of these relations
\begin{equation}
2-\a=3\nu\,,\qquad \gamma=\nu(2-\eta)\,,
\eqlabel{use1}
\end{equation}
and \eqref{sumcrit}, we find
\begin{equation}
\{\nu,\eta\}\bigg|_{scaling}=\left\{\frac12, 1\right\}\,.
\eqlabel{use2}
\end{equation}
Much like in \cite{cc3}, the non-vanishing of the anomalous critical exponent $\eta$ conflicts with the expectation
that for large-N gauge theory plasmas 
the continuous phase transitions in holographic models should be of mean-field type $\eta_{mean-field}=0$.

A relaxation of the system to equilibrium in the vicinity of the critical point is 
commonly discussed within the theory of the dynamical critical phenomena developed by 
Hohenberg and Halperin \cite{hh}. According to \cite{hh} a model is designated to a 
specific universality class based on the dimensionality, symmetries of the order parameter, 
the presence of any conserved densities, and any other properties that affect the 
static critical behavior. A representative of a given dynamical universality class   
is then characterized by a {\it dynamical} critical exponent $z$. This critical exponent 
determines the scaling of the non-equilibrium (time-dependent)  two-point 
correlation function of the order parameter (magnetization in our case) 
at criticality, \ie
\begin{equation}
\tilde{G}(\omega,\vec{q},t,\calh)=\l^{2 y_{\calh}-3+z}
\tilde{G}(\l^z\omega,\l\vec{q},\l^{y_T}t,\l^{y_\calh}\calh)\,,
\eqlabel{dynsc1}
\end{equation}  
for its space-time Fourier transform. The equilibration of a dynamical system is thus
characterized by a relaxation time $\tau$
\begin{equation}
\tau\propto \xi^z\,,
\eqlabel{taudef}
\end{equation} 
which (for $z\ne 0$) diverges at criticality.
The absence of any conserved order parameters puts  $\caln=2^*$ plasma in the 
universality class of `model A'  according to the classification of Hohenberg and Halperin, 
and predicts 
\begin{equation}
z\bigg|_{prediction}=2+c\eta\,,
\eqlabel{zpred}
\end{equation}  
where the constant $c$ can be computed via renormalization group calculations in 
$p=4-\epsilon\,, \epsilon\ll 1\,,$ spatial dimensions, and $\eta$ is the anomalous critical exponent.

In the rest of this section we introduce dynamical susceptibility 
of $\caln=2^*$ plasma and explain how is can be used to 
compute the static critical exponents $\{\nu,\eta\}$ and the dynamical 
critical exponent $z$ of its phase transition. 

\subsection{Dynamical susceptibility of $\caln=2^*$ plasma --- gauge theory perspective}

Both the  critical exponents $\{\nu,\eta\}$ and the dynamical exponent 
$z$ can be extracted from the dynamical
susceptibility of the system.
Consider the response of the system to the time-dependent inhomogeneous 
variations of the external magnetic field $\calh$,
\begin{equation}
\calh\to \calh+\dd\calh(t,\vec{x})\,,\qquad \dd\calh=\int \frac{d^3k}{(2\pi)^3}\ 
\int \frac{d\omega}{2\pi}\
e^{i\vec{k}\cdot \vec{x}-i\omega t}\ \calh_{\omega,\vec{k}}\,.
\eqlabel{delh}
\end{equation}
At the linearized level the variation of the external magnetic field would 
produce a corresponding variation in the magnetization $\dd\calm(t,\vec{x})$
( $\calm_{\omega,\vec{k}}$ for the Fourier components). Following \cite{hh} 
we introduce the dynamical susceptibility as 
\begin{equation}
\chi_{\omega,\vec{k}}=\left(\frac{\calm_{\omega,\vec{k}}}{\calh_{\omega,\vec{k}}}\right)\bigg|_T
\,,\qquad \lim_{(\omega,\vec{k})\to (0,\vec{0})}\ \chi_{\omega,\vec{k}}=\chi_T=\left(\frac{\del\calm}
{\del\calh}\right)\bigg|_T\,.
\eqlabel{defst}
\end{equation}
By the equipartition theorem, the static susceptibility 
\begin{equation}
\chi_{\vec{k}}\equiv \chi_{\omega=0,\vec{k}}\,,
\eqlabel{stsus}
\end{equation}
is related to the Fourier transform $\tilde{G}(\vec{k})$
of the magnetization variation two-point correlation function
\begin{equation}
G(\vec{x})=\langle\dd\calm(\vec{x})\dd\calm(\vec{0})\rangle_{\dd\calh=0}\,,
\eqlabel{2pt}
\end{equation}  
as 
\begin{equation}
\tilde{G}(\vec{k})=T\chi_{\vec{k}}\,.
\eqlabel{schi}
\end{equation}
Given the near-critical behavior of the correlation 
function \eqref{2pt} (see \eqref{defs01}), \eqref{schi} implies that the static susceptibility 
$\chi_{\vec{k}}$ has a pole at 
\begin{equation}
k^2\propto -\xi^{-2}\,,
\eqlabel{polecrit}
\end{equation}
in the vicinity, but not right at the critical point. 
On the other hand, right at the critical point 
\begin{equation}
\chi_{\vec{k}}\propto |\vec{k}|^{-2+\eta}\,.
\eqlabel{critsus}
\end{equation}

The theory of dynamical critical phenomena \cite{hh} predicts that in the vicinity of the continuous phase transition,
and for $|\vec{k}|\sim \xi^{-1}$ the full dynamical susceptibility $\chi_{\omega,\vec{k}}$
will develop a pole at 
\begin{equation}
\omega\sim -i\xi^{-z}\,,
\eqlabel{dyn1}
\end{equation}
with $z$ being the dynamical critical exponent of the system. The frequency in \eqref{dyn1}
(in the hydrodynamic limit) defines a relaxation time $\tau$ as 
\begin{equation}
\t^{-1}\equiv i\omega\propto \xi^{-z}\,.
\eqlabel{dyn2}
\end{equation}

To summarize, following the position of the poles in the static susceptibility $\chi_{\vec{k}}$
as a function of the reduced temperature $t\ne 0$
\begin{equation}
0=\frac{1}{\chi_{\vec{k}}}\ \bigg|_{|\vec{k}|^2=k_*^2(t)}\qquad \ra\qquad k_*^2(t)\propto -\xi^{-2}\propto -t^{2\nu}\,,
\eqlabel{statnu}
\end{equation}
would determine the critical exponent $\nu$; the critical exponent $\eta$ is determined from the static susceptibility 
scaling at critical temperature, \ie $t=0$, as in \eqref{critsus}.
Likewise, scaling of the pole in the dynamical susceptibility $\chi_{\omega,\vec {k}}$
in the hydrodynamic limit as a function of the reduced temperature $t\ne 0$ determines 
the dynamical critical exponent $z$:
\begin{equation}
0=\frac{1}{\chi_{\omega,\vec{k}}}\ \bigg|_{(\omega=\omega_*(t), \vec{k}\to \vec{0})}\qquad \ra\qquad i\omega_*(t)\propto t^{-z\nu}\,.
\eqlabel{dynz}
\end{equation}

\subsection{Dynamical susceptibility of $\caln=2^*$ plasma --- gravity perspective}

In case of $\caln=2^*$ plasma we identify the external magnetic field 
$\calh$ with the bosonic mass $m_b$, \eqref{map}. The variation 
$\calh_{\omega,\vec{k}}$  
would correspond to the variation in 
\begin{equation}
m_b\to m_b+\dd m_b(\omega,\vec{k}) e^{i \vec{k}\cdot \vec{x}-i\omega t }\,,
\eqlabel{var1}
\end{equation}
which 
on the gravity side can be induced by the variation in the 
non-normalizable coefficient $\r_{11}$ the supergravity scalar $\r$:
\begin{equation}
\r_{11}\to \r_{11}+\dd\r_{11}(\omega,\vec{k}) e^{i \vec{k}\cdot \vec{x}-i\omega t }\,,
\qquad  \dd\r_{11}(\omega,\vec{k})\propto 
\dd m_b(\omega,\vec{k})\,. 
\eqlabel{var2}
\end{equation}
The variation $\dd\r_{11}(\omega,\vec{k})$ would produce a linearized response in  
the normalizable coefficient $\r_{10}(\omega,\vec{k})$ 
of the supergravity scalar $\r$:
\begin{equation}
\r_{11}\to \r_{11}+\dd\r_{11}(\omega,\vec{k})
 e^{i \vec{k}\cdot \vec{x}-i\omega t }\qquad \ra\qquad \r_{10}\to \r_{10}
+\dd\r_{10}(\omega,\vec{k}) e^{i \vec{k}\cdot \vec{x}-i\omega t }\,.
\eqlabel{var3}
\end{equation} 
Thus it is natural to identify the variation $\dd\r_{10}(\omega,\vec{k})$ with the 
variation in the magnetization $\calm_{\omega,\vec{k}}$
\begin{equation}
\dd\r_{10}(\omega,\vec{k})\propto \calm_{\omega,\vec{k}}\,.
\eqlabel{var4}
\end{equation}
Finally, the dynamical susceptibility \eqref{defst} is related to the dual 
gravitational data as 
\begin{equation}
\chi_{\omega,\vec{k}}\propto 
\frac{\dd\r_{10}(\omega,\vec{k})}{\dd\r_{11}(\omega,\vec{k})}\,.
\eqlabel{var5}
\end{equation}
The identification \eqref{var5} is equivalent to the one made in 
recent analysis of the holographic critical phenomena \cite{cc1,cc3}.

The holographic computation of the susceptibility \eqref{var5} necessitates the 
analysis of the linearized fluctuations in the gravitational background 
\eqref{background}. The relevant fluctuations were studied previously 
in \cite{bbs,bp}. We briefly review the basic setup of such computations here.
Without the loss of generality we can assume that 
\begin{equation}
k^i=q\ \dd^i_3\,.
\eqlabel{kqdef}
\end{equation}
The linearized on-shell fluctuation of the gravitational 
scalar\footnote{For a critical phenomena with $m_f=0$ we can consistently 
truncate the effective action \eqref{action5} to $\chi=0$.} 
$\a=\ln\r$,
\begin{equation}
\a\to \a+\phi(x)\ e^{i q x_3-i \omega t}\,,
\eqlabel{var6}
\end{equation}
couples to the on-shell fluctuations in the background metric 
\begin{equation}
g_{\mu\nu}\to g_{\mu\nu}+h_{\mu\nu}(x)\ e^{i q x_3-i\omega t}\,,
\eqlabel{var7}
\end{equation}
which form a helicity-0 representation with respect to rotations 
about $x_3$-axis:
\begin{equation}
\left\{h_{tt},\, h_{t x_3},\, h_{aa}=h_{x_1x_1}+h_{x_2x_2},\, h_{x_3x_3}\right\} \,.
\end{equation}
Note that we partially fixed the background diffeomorphisms with 
\begin{equation}
h_{tx}=h_{x_3 x}=h_{xx}=0\,.
\eqlabel{diffeofixed}
\end{equation} 
Following \cite{bbs} we introduce the diffeomorphism-invariant 
linear combinations of fluctuations 
\begin{equation}
Z_H=4\frac{q}{\omega} H_{tz}+2 H_{zz}-H_{aa}\left(1+\frac{q^2}{\omega^2}
\frac{g_{tt}'}{g_{x_1x_1}'}\right)+2 \frac{q^2}{\omega^2}(1-x)^2 H_{tt}\,,
\eqlabel{zhdef}
\end{equation}
\begin{equation}
Z_\phi=\phi-\frac{\a'}{2(\ln g_{x_1x_1})'}H_{aa} \,,
\eqlabel{zadef}
\end{equation}
where $g_{tt}(x)$ and $g_{x_1x_1}(x)$ are the corresponding 
components of the background metric \eqref{background}, the derivatives 
are with respect to $x$ and 
\begin{equation}
h_{tt}=-g_{tt} H_{tt}\,,\qquad h_{tz}=g_{x_1x_1} H_{tz}\,,\qquad 
h_{aa}=g_{x_1x_1} H_{aa}\,,\qquad h_{x_3x_3}=g_{x_1x_1} H_{x_3x_3}\,.
\eqlabel{defH}
\end{equation}
Introduce 
\begin{equation}
\ww\equiv\frac{\omega}{2\pi T}\,,\qquad \qq\equiv\frac{q}{2\pi T}\,.
\eqlabel{wq}
\end{equation}
The equations of motion for $\{Z_H,Z_\phi\}$ take the form
\begin{equation}
\begin{split}
0=&Z_H''+\calc_{11}\ Z_H'+\calc_{12}\ Z_\phi'+\calc_{13}\ Z_H+\calc_{14}\ Z_\phi\,,\\
0=&Z_\phi''+\calc_{21}\ Z_H'+\calc_{22}\ Z_\phi'+\calc_{23}\ Z_H+\calc_{24}\ Z_\phi\,,
\end{split}
\eqlabel{zz}
\end{equation}  
where the  coefficients $\calc_{ij}$ are nonlinear functionals of 
the background fields $\{\r,a\}$ with explicit dependence on $x$ 
and $\{\ww,\qq\}$ \cite{bbs}: 
\begin{equation}
\calc_{ij}=\calc_{ij}\bigg[\{\r,a\};\, x;\, \{\ww,\qq\}\bigg]\,.
\eqlabel{cij}
\end{equation}
Since the equations \eqref{zz} are homogeneous, we can always set 
the non-normalizable component of $Z_\phi$ --- which is the 
diffeomorphism-invariance analog of $\dd\r_{11}(\omega,\vec{k})$ --- 
to one; the dynamical susceptibility is then proportional to the 
normalizable component of $Z_\phi$.

We can summarize now the boundary value problem whose solution would determine the dynamical susceptibility.
Introducing\footnote{The $\ww-$ and $\qq-$dependent rescaling are for convenience in further analysis.} 
\begin{equation}
\begin{split}
Z_H=&(1-x)^{-i\ww}\ \ww^{-2}\ z_H(x,\ww,\qq)\,,\\
Z_\phi=&(1-x)^{-i\ww}\ \qq^{-2}\ z_\phi(x,\ww,\qq)\,,
\end{split}
\eqlabel{expsus}
\end{equation}
the equations of motion for $\{z_H,z_\phi\}$ are solved with the following boundary conditions:
\begin{equation}
\begin{split}
&\lim_{x\to 1_-}z_H=\lim_{x\to 1_-}z_\phi={\rm finite}\,,\\
&z_H=\calo(x)\,,\qquad  z_\phi=(\ln x+\calz(\ww,\qq)) x^{1/2}+
\calo(x\ln^2 x)\,,\qquad {\rm as}\ x\to 0_+\,.
\end{split}
\eqlabel{ssbc}
\end{equation}
The normalizable component $\calz$ of $z_\phi$ near the boundary  is proportional to the dynamical 
susceptibility:
\begin{equation}
\chi_{\omega,\vec{k}}\equiv \chi_{\ww,\qq}\propto \calz(\ww,\qq) \,.
\eqlabel{czrel}
\end{equation}

\section{Critical exponents $\{\nu,\eta\}$ and $z$ of $\caln=2^*$ plasma}

\begin{figure}[t]
\begin{center}
\psfrag{zs}{{$\calz^{-1}\propto \chi_{\ww=0,\qq=0}^{-1}$}}
\psfrag{r11}{{$\r_{11}$}}
\includegraphics[width=4in]{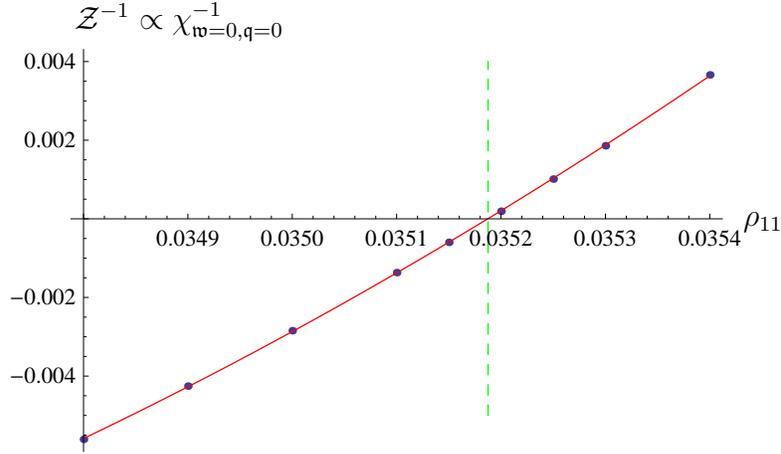}
\end{center}
  \caption{(Colour online) The scaling (blue dots) of the inverse of the 
static susceptibility $\chi_{\ww=0,\qq=0}$ in the vicinity of the critical 
point. The solid red line is a quadratic  fit to data, the dashed green
line represents $\r_{11}=\r_{11}^{c}$. 
} \label{figure3}
\end{figure}

\begin{figure}[t]
\begin{center}
\psfrag{zs}{{$\qq_*^2$}}
\psfrag{r11}{{$\r_{11}$}}
\includegraphics[width=4in]{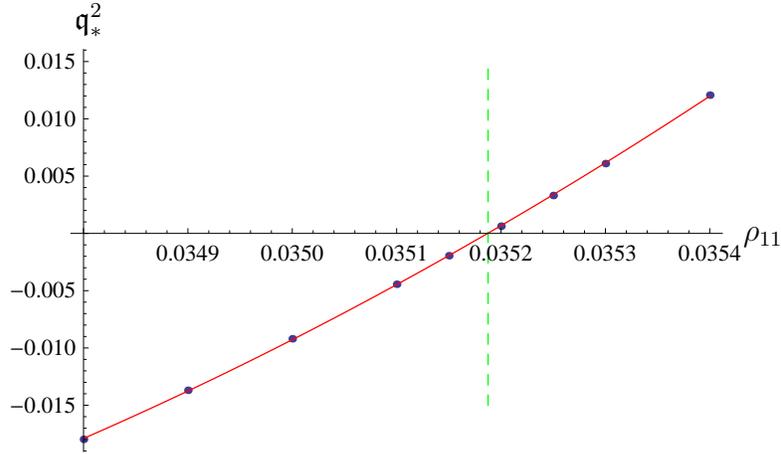}
\end{center}
  \caption{(Colour online) Poles of the static susceptibility in 
the vicinity of the critical point: $\chi^{-1}_{\ww=0,\qq=\qq_*}=0$. 
The solid red line is a quadratic fit to data, the dashed green
line represents $\r_{11}=\r_{11}^{c}$. 
} \label{figure4}
\end{figure}

\begin{figure}[t]
\begin{center}
\psfrag{qs2}{{$\qq^2$}}
\psfrag{zc}{{$(\calz^{crit})^{-1}\propto (\c_{c,\ww=0,\qq}^{crit})^{-1}$}}
  \includegraphics[width=4in]{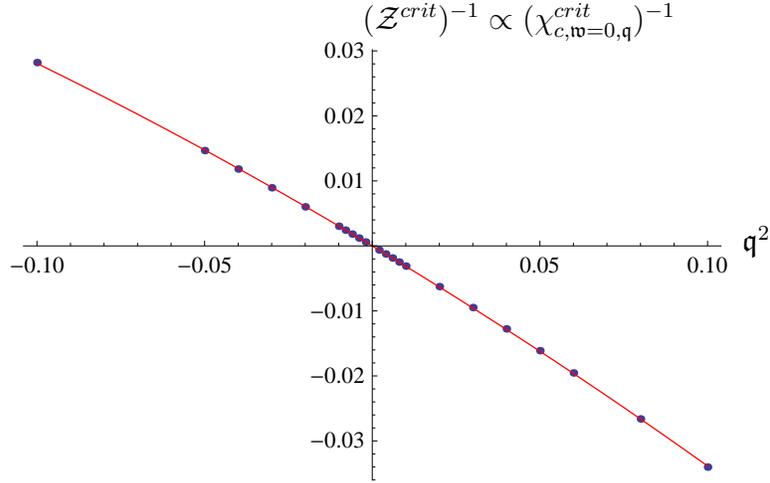}
\end{center}
  \caption{(Colour online) The scaling (blue dots) of the inverse of the static susceptibility 
$\c_{\ww=0,\qq}^{crit}$ at  the critical point, $\r_{11}=\r_{11}^c$. The solid red line is a quadratic fit to the data. 
} \label{figure5}
\end{figure}

\begin{figure}[t]
\begin{center}
\psfrag{iw}{{$i\ww$}}
\psfrag{chi}{{$\calz^{-1}\propto \c^{-1}_{\ww,\qq=10^{-2}}$}}
\includegraphics[width=3in]{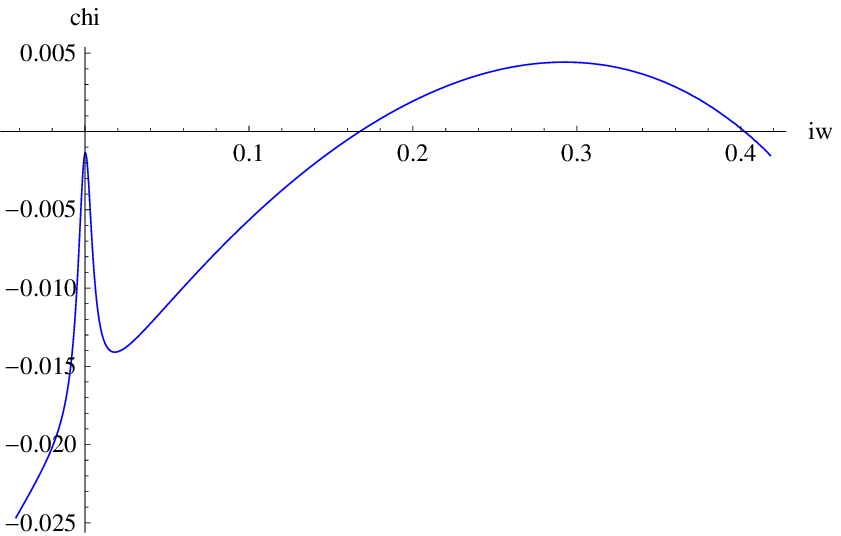}
  \includegraphics[width=3in]{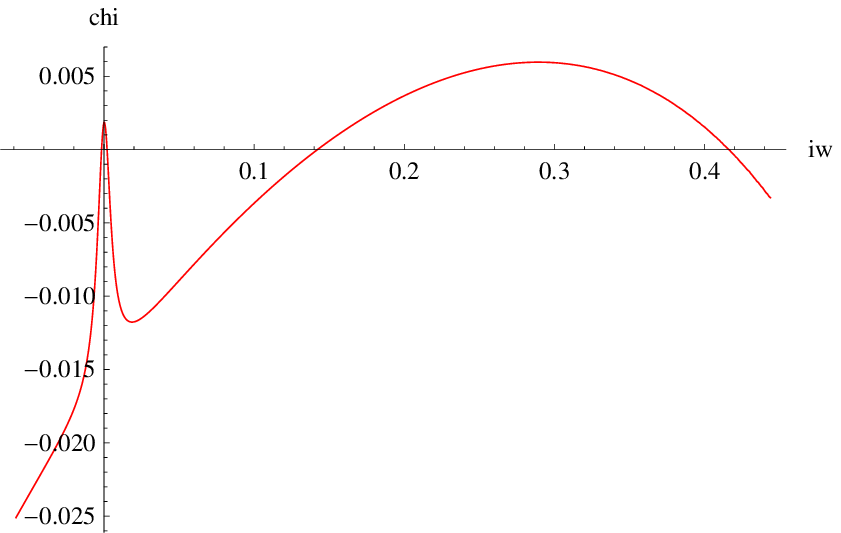}
\end{center}
  \caption{(Colour online)
The inverse of the dynamical susceptibility of $\caln=2^*$
plasma in a stable phase (blue curve) and an unstable phase 
(red curve) at $\qq=10^{-2}$.} \label{figure6}
\end{figure}

\begin{figure}[t]
\begin{center}
\psfrag{iw}{{$\frac{i\ww}{\qq}$}}
\psfrag{chi}{{$\calz^{-1}\propto \c^{-1}_{\ww,\qq=10^{-2}}$}}
\includegraphics[width=3in]{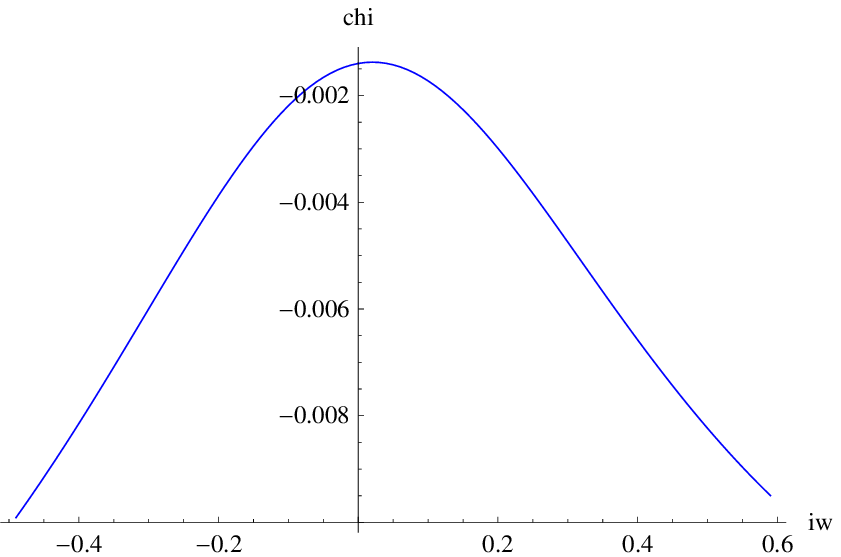}
  \includegraphics[width=3in]{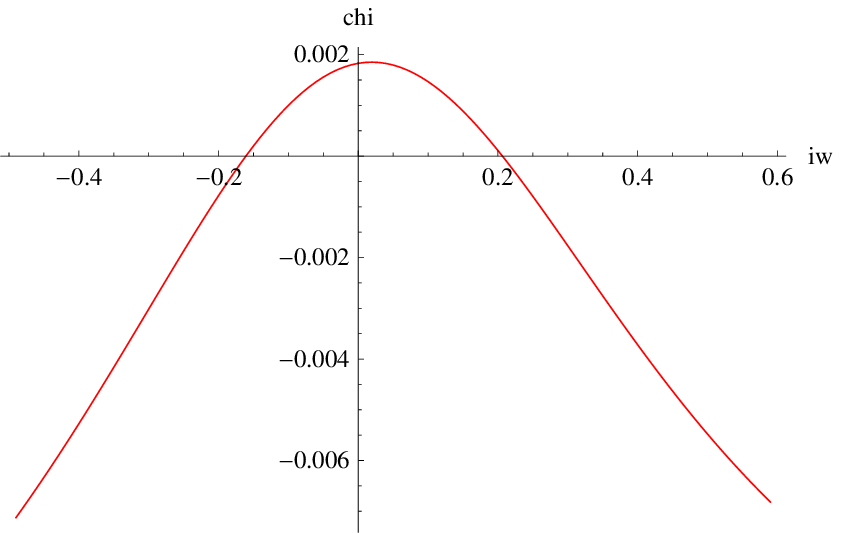}
\end{center}
  \caption{(Colour online)
The inverse of the dynamical susceptibility of $\caln=2^*$
plasma in a stable phase (blue curve) and an unstable phase 
(red curve) at $\qq=10^{-2}$ for $|i\ww|\ll1$.} \label{figure7}
\end{figure}

\begin{figure}[t]
\begin{center}
\psfrag{iw}{{$i\ww_{*,L}\ \&\ i\ww_{*,R}$}}
\psfrag{r11}{{$\r_{11}$}}
\includegraphics[width=4in]{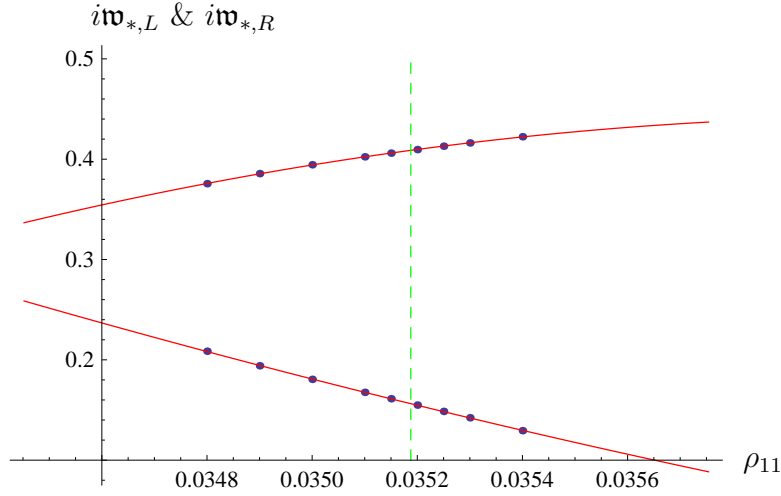}
\end{center}
  \caption{(Colour online) The relaxations rates (blue dots) 
$i\ww_{*,\{L,R\}}$ ($\{$bottom,top$\}$) of $\caln=2^*$
plasma in the vicinity of the critical point. The solid red lines are the  
quadratic fits to the data, and the dashed  green line represents $\r_{11}=\r_{11}^c$.
} \label{figure8}
\end{figure}

In previous section we explained how the dynamical susceptibility 
$\chi_{\ww,\qq}$
can be used to extract the static critical exponents $\{\nu,\eta\}$
and the dynamical critical exponent $z$ of a phase transition. 
We also related this susceptibility 
to the normalizable component $\calz(\ww,\qq)$ of the gravitational 
scalar, which non-normalizable component played the role of the 
(time-dependent and inhomogeneous) variation of the external magnetic field,
see \eqref{czrel}. Here, without going into the technical details of the 
analysis of the boundary value problem \eqref{zz}-\eqref{ssbc}, we present 
the results.

Figure \ref{figure3} shows the inverse of the static susceptibility 
at $\qq=0$ (blue dots) in the vicinity of the critical point. The solid 
red line $Z_{fit}^{-1}=Z_{fit}^{-1}(\r_{11})$
is the best quadratic fit to the data. The vertical green line 
denotes the critical value of $\r_{11}$, \ie $\r_{11}^c$ \eqref{derdr},
corresponding to critical temperature $T_c$, see \eqref{mftc}.
The red line intersects the $\r_{11}$ axis at $\r_{11}^*$, such that
\begin{equation}
\calz_{fit}^{-1}\bigg|_{\r_{11}=\r_{11}^*}=0\qquad \ra 
\left|\frac{\r_{11}^*}{\r_{11}^{c}}-1\right|=9.(3)\times 10^{-6} \,,
\eqlabel{rint}
\end{equation}
in excellent agreement with the critical behavior of $\chi_T$ 
deduced from the thermodynamics \eqref{gamma}:
\begin{equation}
\calz^{-1}_{fit}\propto \Delta\r_{11}\propto t^{1/2}\qquad 
\Longleftrightarrow\qquad \chi_T^{-1}\propto t^{1/2}\,. 
\eqlabel{scast}
\end{equation}

Figure \ref{figure4} presents the poles (blue dots) of the static susceptibility at $\qq=\qq_*$
in the vicinity of the critical point:
\begin{equation}
\c_{\ww=0,\qq=\qq_*}^{-1}=0\,.
\eqlabel{statpoles}
\end{equation}
The solid red line represents the best quadratic fit to the data, and the vertical green line 
denotes the critical value of $\r_{11}$, \ie $\r_{11}^c$,  see \eqref{derdr}.
Notice that in the stable phase, \ie for  $\r_{11}< \r_{11}^{c}$, in the vicinity of the 
phase transition the poles in the static susceptibility are for purely imaginary momenta,
which implies the exponential decay of the magnetization density two-point correlation function
\eqref{2pt}. Furthermore, from \eqref{statnu} we identify the correlation length 
as 
\begin{equation}
(2\pi T_c\ \xi)^2\ \propto\ -\qq_*^{-2}\ \propto\ \frac{1}{|\Delta\r_{11}|}\ \propto +t^{-1/2}\,,\qquad 
0<\r_{11}^c-\r_{11}\ll \r_{11}^c\,,
\eqlabel{xiscale}
\end{equation} 
where we used the results of the fit  and the relation between $\r_{11}$  
and the reduced temperature $t$ \eqref{derdr}. From \eqref{xiscale} we extract the 
(static) critical exponent $\nu$: 
\begin{equation}
\xi\ \propto\ t^{-\nu}\ \propto\ t^{-1/4}\qquad \Rightarrow\qquad \nu=\frac 14 \,.
\eqlabel{xiscale2}
\end{equation}
Given that the static critical exponent $\a=\frac 12$, \eqref{xiscale2} implies 
that the hyperscaling relation \eqref{use1} is violated
\begin{equation}
2-\a\ \ne 3\ \nu\,.
\eqlabel{hyperscale}
\end{equation}

Figure \ref{figure5} shows the inverse of the static susceptibility as a function of $\qq$ (blue dots) 
right at the critical point $\r_{11}=\r_{11}^c$. The solid red line represents the best quadratic fit to the data 
\begin{equation}
(\calz^{crit}_{fit})^{-1}=-5.55084\ \cdot 10^{-6} - 0.30963\ \qq^2 -0.28859\ \qq^4+\calo(\qq^6)\,.
\eqlabel{red6}
\end{equation}
 The red line \eqref{red6} 
intersects the $\qq^2$ axis at 
\begin{equation}
\qq^2_c=-1.8\ \cdot 10^{-5}\,,
\eqlabel{qcsus}
\end{equation}
in excellent agreement with the expected value $\qq_c^2=0$ \eqref{critsus}. 
The data implies  
\begin{equation}
\c^{crit}_{\ww=0,\qq}\ \propto\ \calz^{crit}\ \propto \qq^{-2}\qquad \Longleftrightarrow\qquad
  \c^{crit}_{\ww=0,\qq}\ \propto\ 
\qq^{-2+\eta}\,,
\eqlabel{defeta}
\end{equation}
which determines the anomalous critical exponent $\eta$ as 
\begin{equation}
\eta=0\,.
\eqlabel{etares}
\end{equation}

Finally, we turn to the dynamical critical exponent $z$. According to 
\eqref{dynz}, it can be extracted from the scaling of the pole in the 
dynamical susceptibility in the vicinity of the critical point (in the hydrodynamic limit)
\begin{equation}
0=\c^{-1}_{\ww=\ww_*,\qq}\,,\qquad i\ww_*\bigg|_{\qq\sim \xi^{-1}}\propto t^{-z\nu}\,.
\eqlabel{polescaling}
\end{equation}
A typical behavior of  the dynamical susceptibility in a thermodynamically 
stable phase, \ie for $\r_{11}=0.0351<\r_{11}^c$ (blue curve), and a thermodynamically 
unstable phase, \ie for $\r_{11}=0.0353>\r_{11}^c$ (red curve) is presented in 
Figure \ref{figure6}. We used $\qq=10^{-2}$. Notice that in the stable phase, 
dynamical susceptibility has two separate poles 
\begin{equation}
0=\c^{-1}_{\ww=\ww_*,\qq=10^{-2}}\,,
\eqlabel{poless}
\end{equation}
for $i\ww_*>0$ ---  according to \eqref{dyn2} these poles correspond to 
{\it different} relaxation time-scales $\t$ of the system at criticality. 
In the unstable phase two additional poles appear for small values of $i\ww$.
Figure \ref{figure7} zooms in on the range of $|i\ww|\ll 1$ in the dynamical 
susceptibility. One of these poles is at negative values of $i\ww_*$, corresponding 
to a negative relaxation time $\t$. A detailed analysis show\footnote{Available 
from authors upon request.} that the negative relaxation time scales as 
$\t^{-1}\equiv i\ww_*\propto -\qq$. Clearly, negative relaxation times 
signal the instability in the system --- this is precisely the instability 
due to the hydrodynamic (sound-channel) modes which propagate with $c_s^2\le 0$
once $\r_{11}\ge \r_{11}^c$, see Figure \ref{figure1}. Such instabilities are expected on
general grounds: whenever a thermodynamic phase of a system has a negative specific 
heat, the hydrodynamic modes in the system are unstable 
\cite{bins}\footnote{The reverse is not true: a
thermodynamically stable system might still have instabilities \cite{bpt}.}. 
What we demonstrated here is that such instabilities 
also result in the negative relaxation time: \ie instead of approaching the 
equilibrium a system is driven away from it. 
 
We now focus on  poles in the dynamical susceptibility at $i\ww>0$:
\begin{equation}
0=\c^{-1}_{\ww,\qq}\bigg|_{\ww=\{\ww_{*,L}(\qq)\,,\ \ww_{*,R}(\qq)\}}\,,
\eqlabel{polesposi}
\end{equation} 
where $L$ and $R$ are the indexes of the two positive poles, such that 
the corresponding relaxation rates
\begin{equation} 
(2\pi T\t_{L}(\qq))^{-1}\equiv i\ww_{*,L}(\qq)\ <\ (2\pi T \t_{R}(\qq))^{-1}\equiv i\ww_{*,R}(\qq)\,.
\eqlabel{inequ}
\end{equation}
We performed numerical analysis for different values  $\qq=\{10^{-4},10^{-3},10^{-2}\}$ ---
the poles $i\ww_{*,\{L,R\}}(\qq)$ have a well defined hydrodynamic limit $\qq\to 0$, which 
is obtained by computing the susceptibilities ( at different values of $\r_{11}$ ) strictly at 
$\qq=0$: 
\begin{equation}
\lim_{\qq\to 0}\ i\ww_{*,\{L,R\}}(\qq) = i\ww_{*,\{L,R\}}(0)\equiv i\ww_{*,\{L,R\}}\,.
\end{equation}
The results of such analysis are presented in Figure 8. 
The blue dots are the relaxation rates of $\caln=2^*$ plasma at criticality, 
the solid red lines are the best quadratic fits to the data. The top dots/curve
corresponds to $i\ww_{*,R}$, and the bottom dots/curve 
corresponds to $i\ww_{*,L}$. Once again, the vertical green dashed line corresponds to 
$\r_{11}=\r_{11}^c$. The results of the analysis show that in the hydrodynamic limit,
both the relaxation rates are finite
\begin{equation}
(2\pi T_c\ \t_{\{L,R\}})^{-1}=i\ww_{*,\{L,R\}}\ \propto\ (\Delta\r_{11})^0\ \propto\ t^0\
\propto\ (2\pi T_c\ \xi)^0\,.
\eqlabel{z}
\end{equation} 
Thus, 
\begin{equation}
\t_{\{L,R\}}\ \propto \xi^z\ \propto\ \xi^0 \qquad \ra\qquad z=0\,.
\eqlabel{zfin}
\end{equation}

In the previous section we pointed out that the critical behavior of $\caln=2^*$ 
plasma should be identified with  that of 'model A' according to dynamical critical 
phenomena classification in \cite{hh}. As such, the dynamical critical exponent $z$ 
was predicted to be \eqref{zpred}
\begin{equation}
z\bigg|_{prediction}=2+c\cdot 0=2\,,
\eqlabel{zpredf}
\end{equation}
which differs from the value we obtained \eqref{zfin}.

\section{Conclusions} 

In this paper, building up on the previous work \cite{cc1,cc2,cc3},
we presented a detailed analysis of the static and dynamic critical 
phenomena in strongly coupled $\caln=2^*$ plasma.  This model 
is a string theory derived example of gauge theory/gravity correspondence 
where one deforms $\caln=4$ SYM by giving a mass $m_b$ to bosonic components
and a mass $m_f$ to fermionic components 
of $\caln=2$ hypermultiplet. Generically, \ie when $m_b\ne m_f$,
such a  deformation completely breaks the supersymmetry. At finite temperature 
$\caln=2^*$ gauge theory plasma undergoes a second-order phase transition, 
provided $m_f^2<m_b^2$. This continuous transition is characterized by
a terminal temperature $T_c$,  which can be reached within isotropic and homogeneous 
equilibrium phases. At temperature $T_c$ the two phases continuously meet, with vanishing 
speed of sound. One of the phases is always perturbatively unstable --- the instabilities 
reside in the sound-channel hydrodynamic modes which propagate with $c_s^2<0$. 
Extending \cite{cc2}, we computed the static and the dynamical critical exponents 
of the transition, as approached from the perturbatively stable phase:
\begin{equation}
\left(\a,\b,\gamma,\delta,\nu,\eta;\ z\right)=\left(\frac 12,\frac 12,\frac 12, 2,\frac 14,0;\
0 \right)\,.
\eqlabel{exponents}
\end{equation}     
As expected --- since the gauge theory is a large-$N$ model ---
 the transition is of the mean-field theory type with vanishing anomalous static critical
exponent $\eta$. Under the static scaling hypothesis, only two of the critical exponents 
are independent --- thus, the six static critical exponents in \eqref{exponents} 
must satisfy four algebraic constraints. Similar to analysis in \cite{cc3}, 
we find that only one of these constraints, \ie the hyperscaling relation, is being violated
\begin{equation}
2-\a\ne 3\nu\,.
\eqlabel{violation}
\end{equation} 

Dynamical features of the transition in $\caln=2^*$ plasma are quite interesting. 
First of all, the symmetries of the transition identify it as the one in the universality 
class of 'model A', according to classification of Hohenberg and Halperin \cite{hh}. 
The latter predicts the dynamical critical exponent as $z_{prediction}=2$, which contradicts 
direct computations \eqref{exponents}. Both the stable and the unstable phases 
have multiple (two) relaxation times which remain finite at the critical point ---
hence the critical exponent $z=0$. Once again, as in analysis in \cite{cc2},
even though the dynamical critical exponent $z\ne 1$, and thus there is 
an anisotropy between the time- and the space- coordinates scaling 
in the two-point (non-equilibrium) correlation functions, the dual gravitational 
geometry at criticality does not exhibit a Lifshitz-like scaling in the 
sense of \cite{klm}. In fact, if, as suggested by \cite{cc3}, 
different non-equilibrium correlation functions at criticality have different 
dynamical exponents $z$, it is not possible to 'by hand' embed the anisotropic
scaling of the correlations functions into the symmetric of the dual geometry.
The critical phenomena in \cite{cc3} and the one considered here indicates 
that anisotropic scaling is rather an emergent phenomena.  
Second, the unstable phase of $\caln=2^*$ plasma has an additional 
relaxation rate $\t^{-1}_{unstable}\propto -|\vec{k}|$. A negative relaxation rate 
indicates that rather than approaching the equilibrium, a perturbed system is driven 
away from it --- this is yet another reflection of the instability in the hydrodynamic 
sector of the theory, which is necessarily linked to a thermodynamic instability 
of the corresponding plasma phase \cite{bins}.   
  
In the future, it  would be interesting to understand how the general classification 
of dynamical critical phenomena \cite{hh} should be enlarged to incorporate 
the universality class of $\caln=2^*$ plasma. Probably the most pressing question is the 
understanding of the equilibrium phases in this universality class (and also the one 
of the $\caln=4$ SYM plasma with an R-symmetry chemical potential) for $T< T_c$. 
Such phases can not be homogeneous and isotropic.

\section*{Acknowledgments}
Research at Perimeter Institute is
supported by the Government of Canada through Industry Canada and by
the Province of Ontario through the Ministry of Research \&
Innovation. AB gratefully acknowledges further support by an NSERC
Discovery grant and support through the Early Researcher Award
program by the Province of Ontario.


\begin{thebibliography}{99}


\bibitem{m9711}
J.~M.~Maldacena,
Adv.\ Theor.\ Math.\ Phys.\  {\bf 2}, 231 (1998)
[Int.\ J.\ Theor.\ Phys.\  {\bf 38}, 1113 (1999)]
[arXiv:hep-th/9711200].

\bibitem{pw}
  K.~Pilch and N.~P.~Warner,
  Nucl.\ Phys.\  B {\bf 594}, 209 (2001)
  [arXiv:hep-th/0004063].

\bibitem{bpp}
  A.~Buchel, A.~W.~Peet and J.~Polchinski,
  Phys.\ Rev.\ D {\bf 63}, 044009 (2001)
  [arXiv:hep-th/0008076].


\bibitem{ejp}
  N.~J.~Evans, C.~V.~Johnson and M.~Petrini,
  JHEP {\bf 0010}, 022 (2000)
  [arXiv:hep-th/0008081].



\bibitem{u1}
  A.~Buchel and J.~T.~Liu,
  Phys.\ Rev.\ Lett.\  {\bf 93}, 090602 (2004)
  [arXiv:hep-th/0311175].

\bibitem{u2}
  P.~Kovtun, D.~T.~Son and A.~O.~Starinets,
  Phys.\ Rev.\ Lett.\  {\bf 94}, 111601 (2005)
  [arXiv:hep-th/0405231].


\bibitem{u3}
  A.~Buchel,
  Phys.\ Lett.\  B {\bf 609}, 392 (2005)
  [arXiv:hep-th/0408095].

\bibitem{u4}
  P.~Benincasa, A.~Buchel and R.~Naryshkin,
  Phys.\ Lett.\  B {\bf 645}, 309 (2007)
  [arXiv:hep-th/0610145].


\bibitem{bulk1}
  A.~Buchel,
  Phys.\ Lett.\  B {\bf 663}, 286 (2008)
  [arXiv:0708.3459 [hep-th]].

\bibitem{v1}
  Y.~Kats and P.~Petrov,
  JHEP {\bf 0901}, 044 (2009)
  [arXiv:0712.0743 [hep-th]].

\bibitem{v2}
  M.~Brigante, H.~Liu, R.~C.~Myers, S.~Shenker and S.~Yaida,
  Phys.\ Rev.\  D {\bf 77}, 126006 (2008)
  [arXiv:0712.0805 [hep-th]].

\bibitem{v3}
  A.~Buchel, R.~C.~Myers and A.~Sinha,
  JHEP {\bf 0903}, 084 (2009)
  [arXiv:0812.2521 [hep-th]].

\bibitem{gpr}
  S.~S.~Gubser, S.~S.~Pufu and F.~D.~Rocha,
  JHEP {\bf 0808}, 085 (2008)
  [arXiv:0806.0407 [hep-th]].

\bibitem{cc1}
  K.~Maeda, M.~Natsuume and T.~Okamura,
  Phys.\ Rev.\  D {\bf 78}, 106007 (2008)
  [arXiv:0809.4074 [hep-th]].


\bibitem{cc2}
  A.~Buchel and C.~Pagnutti,
  Nucl.\ Phys.\  B {\bf 834}, 222 (2010)
  [arXiv:0912.3212 [hep-th]].

\bibitem{cc3}
  A.~Buchel,
  Nucl.\ Phys.\  B {\bf 841}, 59 (2010)
  [arXiv:1005.0819 [hep-th]].

\bibitem{bl}
  A.~Buchel and J.~T.~Liu,
  JHEP {\bf 0311}, 031 (2003)
  [arXiv:hep-th/0305064].

\bibitem{n2hydro}
  A.~Buchel,
  Nucl.\ Phys.\  B {\bf 708}, 451 (2005)
  [arXiv:hep-th/0406200].

\bibitem{bdkl}
  A.~Buchel, S.~Deakin, P.~Kerner and J.~T.~Liu,
  Nucl.\ Phys.\  B {\bf 784}, 72 (2007)
  [arXiv:hep-th/0701142].

\bibitem{bp}
  A.~Buchel and C.~Pagnutti,
  Nucl.\ Phys.\  B {\bf 816}, 62 (2009)
  [arXiv:0812.3623 [hep-th]].

\bibitem{franco}
  S.~Franco, A.~M.~Garcia-Garcia and D.~Rodriguez-Gomez,
  Phys.\ Rev.\  D {\bf 81}, 041901 (2010)
  [arXiv:0911.1354 [hep-th]].



\bibitem{cca}
  A.~Buchel,
  Nucl.\ Phys.\  B {\bf 820}, 385 (2009)
  [arXiv:0903.3605 [hep-th]].

\bibitem{mr}
  O.~Aharony, S.~S.~Gubser, J.~M.~Maldacena, H.~Ooguri and Y.~Oz,
  Phys.\ Rept.\  {\bf 323}, 183 (2000)
  [arXiv:hep-th/9905111].

\bibitem{abh}
O.~Aharony, A.~Buchel and M.~Heller, work in progress. 

\bibitem{hh}
  P.~C.~Hohenberg and B.~I.~Halperin,
  Rev.\ Mod.\ Phys.\  {\bf 49}, 435 (1977).


\bibitem{bbs}
P.~Benincasa, A.~Buchel and A.~O.~Starinets,
  Nucl.\ Phys.\ B {\bf 733}, 160 (2006)
  [arXiv:hep-th/0507026].
 
\bibitem{bins}
  A.~Buchel,
  Nucl.\ Phys.\  B {\bf 731}, 109 (2005)
  [arXiv:hep-th/0507275].


\bibitem{bpt}
 A.~Buchel and C.~Pagnutti, to appear.

\bibitem{klm}
  S.~Kachru, X.~Liu and M.~Mulligan,
  Phys.\ Rev.\  D {\bf 78}, 106005 (2008)
  [arXiv:0808.1725 [hep-th]].



\end{thebibliography}
\end{document}